\documentstyle[aps,graphicx,floats,prb,color]{revtex}

\def\u2{$\langle u^{2} \rangle$}
\def\lcmo{La$_{1-x}$Ca$_{x}$MnO$_3$}

\begin{document}
\draft
\wideabs{

\title{
Evidence for charge localization in the ferromagnetic phase of 
La$\bf _{1-x}$Ca$\bf _{x}$MnO$\bf _{3}$ from
High real-space-resolution x-ray diffraction 
}

\author { S. J. L. Billinge, Th. Proffen,  V. Petkov } 
\address{ Department of Physics and Astronomy and Center for
          Fundamental Materials Research,\\
          Michigan State University, East Lansing, Michigan 48824-1116.}
\author { J. L. Sarrao }
\address{Los Alamos National Laboratory, Los Alamos, New Mexico 87545.} 
\author { S. Kycia }
\address{Cornell High Energy Synchrotron Source, Ithaca, New York 14853. }

\date{July 21, 1999}

\maketitle


\begin{abstract}
High real-space-resolution atomic pair distribution functions 
of La$_{1-x}$Ca$_{x}$MnO$_{3}$ ($x=0.12$, 0.25 and 0.33)
have been measured using high-energy x-ray powder diffraction to study the
size and shape of the MnO$_6$ octahedron as a function of temperature and
doping.  In the paramagnetic insulating phase we find evidence for 
three distinct bond-lengths ($\sim 1.88$, 1.95 and 2.15~\AA ) which
we ascribe to Mn$^{4+}$-O, Mn$^{3+}$-O short and Mn$^{3+}$-O long bonds 
respectively.
In the ferromagnetic metallic (FM) phase, for $x=0.33$ and $T=20$~K, we 
find a single Mn-O bond-length;
however, as the metal-insulator transition is approached either by increasing
$T$ or decreasing $x$, intensity progressively 
appears around $r=2.15$ and in the region $1.8 - 1.9$~\AA\ 
suggesting the appearance of 
Mn$^{3+}$-O long bonds and short Mn$^{4+}$-O bonds.
This is strong evidence that charge localized and delocalized phases coexist
close to the metal-insulator transition in the FM phase. 
\end{abstract}

}


\section{Introduction}

The importance of the lattice to the colossal magnetoresistance (CMR) 
phenomenon\cite{ramir;jpcm97} is now fairly well 
established.\cite{milli;prl95,roder;prl96,billi;prl96,dai;prb96,booth;prb96,%
zhao;n96}  There is a strong electron-lattice coupling due to the
Jahn-Teller effect which affects Mn$^{3+}$ 
ions\cite{milli;prl95,fletc;jpc69} and the doped carriers tend to
localize as small polarons at high temperature and low 
doping.\cite{billi;prl96,dai;prb96,booth;prb96,louca;prb97,schif;prl95,%
jonke;p50,jaime;prb96,palst;prb97,jaime;prl97,matl;prb98}
However, exact
agreement about the detailed nature of local Jahn-Teller (JT) and
polaronic distortions is lacking.  This information is important for
separating competing models describing the CMR phenomenon.

Early diffraction,\cite{dai;prb96} atomic pair distribution 
function\cite{billi;prl96} (PDF) and extended 
x-ray absorption fine structure\cite{booth;prb96,booth;prb98} (XAFS) studies 
demonstrated that atomic 
disorder, measured as the Mn-O bond-length distribution, increases as 
samples pass through the metal-insulator (MI) transition with temperature.
This is qualitatively what is expected if polarons are forming as the
sample enters the insulating phase.
These techniques also agree that the onset of polaron formation is gradual 
with temperature, taking place over a temperature range of 50-100~K below 
the MI transition temperature $T_m$.  
In general, PDF\cite{billi;prl96,louca;prb97,louca;prb99} and 
XAFS\cite{booth;prb96,booth;prb98,lanza;prl98} results suggest
that in CMR materials the local structure is significantly
different from that observed crystallographically.  In particular,
Mn-O$_6$ octahedra can have a significant JT distortion locally
even when globally the average JT distortion is zero or negligible.
Although the local structural studies agree on this point
there is disagreement on the 
amplitude of the distortions, in particular the length of the 
long JT bond.
For instance, Louca {\it et al.}\cite{louca;prb97,louca;prb99} 
propose, based on the observation of 
a persistent negative fluctuation\cite{lcmonote699} in the neutron PDFs of 
La$_{1-x}$Sr$_{x}$MnO$_3$ at around 2.2-2.3~\AA , that
this is the length of the JT long-bond.  This seems surprising given
that the JT long-bond in the undoped material is shorter at 
2.18~\AA\cite{proff;prb99;unpub}.
On the other hand, XAFS measurements of the Ca doped system\cite{booth;prb96} 
suggest that 
the JT long-bond is between 2.1- 2.2~\AA\ and a difference modeling of 
the neutron PDF from the La$_{1-x}$Ca$_{x}$MnO$_{3}$ system\cite{billi;prl96} 
supports these findings, as we discuss below. 

Another question which is not resolved is the nature of the 
charge ground-state of the ferromagnetic metallic (FM) phase.
Local density approximation
calculations suggest that a delocalized charge state
would not have any JT distortion even when the $e_g$ 
band is not completely empty.  Thus, the ferromagnetic metallic
state, which we refer
to as the Zener state (following Radaelli\cite{radae;mrssp99}),
would have regular undistorted
octahedra.  The observation of essentially undistorted MnO$_6$
octahedra at low temperature in the FM phase is supported by
XAFS\cite{booth;prb98} and PDF results at high enough doping (away
from the low-temperature MI transition).  However, there has
been a prediction based on XAFS data that small octahedral
distortions persist at low temperature in the FM phase suggesting
that the ground-state is a large polaron state.\cite{lanza;prl98} 
PDF data have also
been interpreted in terms of a three-site polaron model\cite{louca;prb97} 
at low temperature persisting at least up to a doping level of $x=0.3$.
It is important to determine the ground state of the FM phase.

Other interesting phenomena also take place in the FM phase
when the MI transition is approached as 
a function of temperature or doping. Upon increase in temperature 
structural distortions start to appear in the local structure below 
$T_c$.\cite{billi;prl96,dai;prb96,booth;prb96}
They also appear when, at low
temperature, doping is decreased towards $x=0.17-0.18$.
In the vicinity of the MI transition 
the FM phase does not seem to be in a pure Zener state. 
The exact nature of this inhomogeneous state is, however, 
not fully characterized.

We have undertaken a high real-space resolution x-ray PDF
study of the \lcmo\ system to try and resolve some of the issues
discussed above. Specifically, by applying the PDF technique we
would like to study the distribution of Mn-O distances in a series of
manganites to elucidate the nature
of the charge ground state of the FM phase: i.e., is it fully delocalized or
not.  We would like to investigate what is the nature of the
local Jahn-Teller and polaronic distortions in this material and
how they evolve as a function of doping and temperature. 
Finally, we would like to differentiate between the competing models
for the evolution of the charge state away from the ground-state
as the MI transition is approached as a function of temperature and
doping. 

By definition, the atomic pair distribution function PDF is the instantaneous
atomic density-density correlation function which describes the
atomic arrangement in materials.\cite{chaik;b;pocmp95} 
It is the sine Fourier transform of the
experimentally observable structure factor obtained in a powder diffraction
experiment.\cite{warre;bk90} Since the total structure function  includes
both the Bragg intensities and diffuse scattering 
its Fourier associate, the PDF, yields both the local
and average atomic structure of materials. By contrast, an analysis of the
Bragg scattering intensities alone yields only the average crystal structure.
Determining the PDF has been the approach of choice for characterizing glasses,
liquids and amorphous solids for a long time.\cite{warre;bk90,wagne;jncs78} 
However, its wide
spread application to crystalline materials, such as manganites, where some
local deviation from the average structure is expected to take place, has
been relatively recent.\cite{egami;b;lsfd98} 

We chose to use high-energy x-rays to measure the PDFs
because it is possible to get high-quality data at high-$Q$ values  
($Q$ is the magnitude of the wavevector) allowing accurate
high real-space resolution PDFs to be determined.\cite{petko;prl99;unpub}
It was previously thought that neutrons were superior for high-$Q$ measurements
because, as a result of the  $Q$-dependence of the x-ray atomic form factor
 the x-ray coherent intensity
gets rather weak at high-$Q$; however, the high-flux of x-rays 
from modern 
synchrotron sources more than compensates for this and we have shown 
that high quality 
high-resolution PDFs can be obtained using x-rays.\cite{petko;prl99;unpub}


\section{Experimental}

The \lcmo\ samples were synthesized by standard solid-state reaction.  
Stoichiometric amounts of La$_{2}$O$_{3}$, CaCO$_{3}$ and MnO$_{2}$ were 
mixed with a mortar and pestle and placed in an alumina crucible. The 
material was fired at 1050~$^\circ$C, 1300~$^\circ$C and 1350~$^\circ$C 
for one day each with 
intermediate grindings.  After the final grinding, the material was 
fired at 1400~$^\circ$C for an additional day and then slow-cooled over 20 hours 
to room temperature.  Samples were characterized by conventional powder  
x-ray diffraction, temperature-dependent magnetization, and electrical 
resistivity.

\begin{figure}[!tb]
  \centering
  \includegraphics[angle=0,width=2.8in]{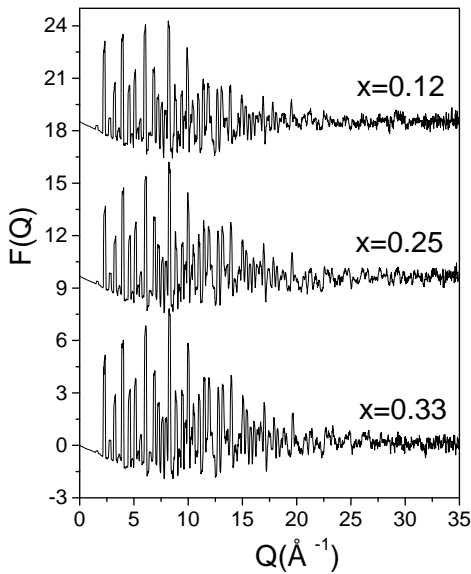}
  \caption{Reduced structure factors $F(Q)=Q[S(Q)-1]$ for  
           \protect\lcmo , $x=0.12$, 0.25 and 0.33, measured
           at 20~K.} 
  \label{fig;soq}
\end{figure}

\begin{figure}[!tb]
  \centering
  \includegraphics[angle=0,width=2.8in]{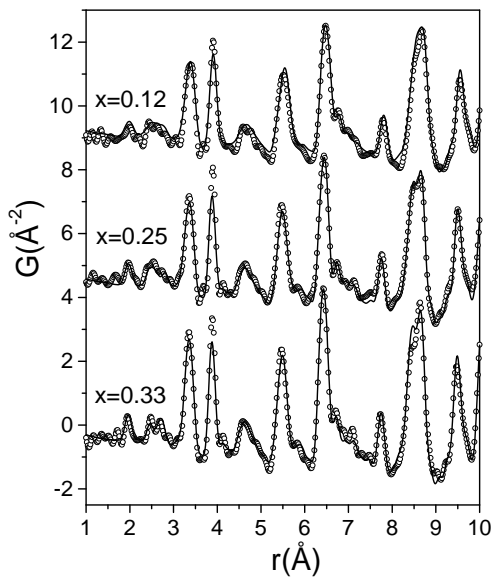}
  \caption{ Reduced radial distribution functions, $G(r)$, for 
           \protect\lcmo , $x=0.12$, 0.25 and 0.33 at 20~K are
           shown as open circles. The corresponding fits
	   are displayed as solid lines (see text for details).} 
  \label{fig;gr}
\end{figure}

Synchrotron powder diffraction experiments were carried out at the A2 56 pole wiggler
beam line at Cornell High Energy Synchrotron Source (CHESS). This
beam line is capable of delivering an intense beam of high energy x-rays
required for high resolution PDF measurements. Data were collected 
in symmetric transmission geometry. The polychromatic incident beam 
was dispersed using a Si (111) monochromator 
and x-rays of energy 61 keV ($\lambda=0.203$\AA) were used. An intrinsic Ge 
detector coupled to a multi-channel analyzer was used to detect the 
scattered radiation, allowing us to extract the coherent component of 
the scattered x-ray intensities by setting appropriate energy windows. 
The diffraction 
spectra were collected by scanning at constant $Q$ steps of 
$\Delta Q=0.025$\AA $^{-1}$. Multiple scans up to 
$Q_{{max}}=40$\AA$^{-1}$ were conducted and the resulting spectra 
averaged to improve the statistical accuracy and reduce any systematic 
error due to instability in the experimental set-up.  
The data were normalized for flux, corrected for background scattering and
experimental effects such as detector deadtime and absorption. The part of
the Compton scattering at low values of Q not eliminated by the preset energy
window was removed analytically applying a procedure suggested by 
Ruland.\cite{rulan;bjap64} The resulting intensities were divided by the 
average atomic form factor for the sample to obtain the total structure 
factor S(Q),
\begin{equation}
  S(Q) = 1 + \frac{I_{c}(Q) - \sum_{i} c_{i}f_{i}^{2}(Q)}
                  {           [\sum_{i} c_{i}f_{i}]^{2}}
\end{equation}
where $I_{c}$ is the measured coherent part of the spectrum, $c_{i}$ and
$f_{i}(Q)$ are the atomic concentration and scattering factor of the atomic
species of type $i$ ($i=$ La, Ca, Mn and O), 
respectively.\cite{wased;bk80} All data 
processing procedures were carried out using the program RAD.\cite{petko;jac89} 
The measured reduced structure factors, $F(Q)=Q[S(Q)-1]$, for $x=0.12$, 
0.25 and 0.33 at $T=20$~K are shown in Figure \ref{fig;soq}.
The data are terminated at $Q_{\mbox{max}} = 35$\AA$^{-1}$
beyond which the signal to noise ratio became unfavorable. 
Note that this a a very high wavevector for x-ray diffraction measurements;
for example, a conventional Cu K$\alpha$ x-ray source has a 
$Q_{max}$ of less than 8~\AA$^{-1}$.
The corresponding reduced atomic distribution functions, $G(r)$, obtained via
Fourier transform
\begin{equation}
  G(r)= \frac{2}{\pi }\int_{0}^{\infty} Q [S(Q)-1] \sin (Qr)\> dQ,
\end{equation}
are shown as open circles in Figure \ref{fig;gr}.
%


\section{Results}

\subsection{Polarons versus JT distortion}

For the sake of clarity we would like to define the terminology
we will use in the following discussion. 
There are two types of octahedral distortions which are observed
in manganites.  The first is a quadrupolar symmetry elongation
of the MnO$_6$  octahedron: i.e., it has two long Mn-O bonds and 
four shorter Mn-O bonds. This is associated with the presence of a
Mn$^{3+}$ ion and is referred to as a Jahn-Teller 
distortion.\cite{sturg;b;ssp67}
Another possible distortion is an isotropic breathing-mode collapse
of the MnO$_6$  octahedron where a regular octahedron stays regular (6 equal bond
lengths) but the octahedron shrinks.  This type of distortion can be
associated with the presence of a Mn$^{4+}$ ion.\cite{billi;prl96}
We refer to this as a polaronic distortion
since the Mn$^{4+}$ ions appear only when doped holes become localized.  

We note that in the literature the Jahn-Teller distorted
octahedra are often referred to as ``Jahn-Teller 
polarons''.\cite{zhao;n96,coey;aip99,polle;jpcs82}  We avoid
this terminology because the presence of Jahn-Teller distorted
octahedra need not imply the presence of polarons in the sense 
of localized doped holes;
for example, the undoped LaMnO$_3$ compound is fully Jahn-Teller
distorted but contains no doped holes.\cite{ellem;jssc71,proff;prb99;unpub}  
Whilst it can be argued that
these are polarons because the Jahn-Teller distortion splits the
$e_g$ band making this compound insulating it confuses the 
discussion of the state, localized or delocalized, of the doped holes.
In our discussion we confine the use of ``polaron'' to describe
a {\it doped hole} localized with an associated lattice distortion.

These doped-hole polarons have also been described in the literature
as ``anti-Jahn-Teller polarons''.\cite{louca;prb99,allen;unpub}  
This terminology comes about when
one considers what happens when a doped hole localizes in a 
background of Jahn-Teller distorted Mn$^{3+}$ octahedra, for
example, in the lightly doped region of the phase diagram. On the 
site where the hole localizes the Jahn-Teller distortion is
locally destroyed; thus the name. 
Again, we avoid this terminology because it is not appropriate
when the polaronic state is approached from the delocalized 
ferromagnetic metallic state (the Zener state).  
In this case, as we discuss below,
there are initially no Jahn-Teller distorted octahedra.  The
octahedra are regular and conform to those seen in the
average crystal structure.  As the
metal-insulator transition is approached the doped holes begin to 
localize.  When they localize both breathing mode collapsed
doped-hole polarons (Mn$^{4+}$) {\it and} Jahn-Teller distorted sites
(Mn$^{3+}$) {\it are created}.
From this perspective it seems confusing to think of the polarons
as ``anti-Jahn-Teller'' polarons. This also raises the point that,
whilst we are not calling the Jahn-Teller distorted octahedra
polarons, in the heavily doped material the presence of fully Jahn-Teller
distorted octahedra implies the presence of localized Mn$^{4+}$
polarons and vice versa.


\subsection{Comparison to the crystal structure}

First we compare the present experimental
PDFs to the average crystal structure determined by
other independent studies.  Experimental PDFs were fit with the 
crystallographic model\cite{kwei;mrs97}
The
refinement was done using the program PDFFIT.\cite{proff;jac99} 
Lattice parameters, isotropic thermal parameters and atomic 
positions were refined conserving the symmetry of the space group ($Pbnm$). 
The
calculated PDFs corresponding to the best fit are shown in Figure~\ref{fig;gr}
 as solid lines. 
Inspection of the figure shows a satisfactory agreement between the
calculated and measured PDFs for all three compositions which shows that the
present experimental PDFs are, in general, consistent with  the average
crystal structure of doped manganites. Furthermore, the values refined reproduce the
Rietveld found values very well. The rather large difference observed for
the PDF peak at $r\approx 4.0$\AA\ is believed to be related to dopant ion
effects on the La/Ca site. Attempts to model these differences are currently
under way. 

\begin{figure}[!tb]
  \centering
  \includegraphics[angle=0,width=2.8in]{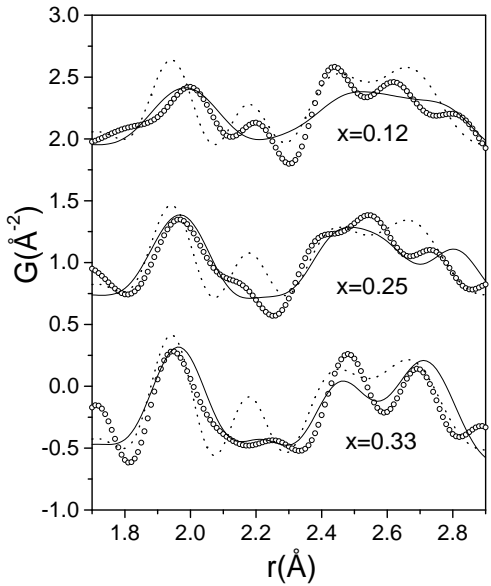}
  \caption{Low $r$ region of the data (open circles) and refinements
           (solid line) shown in Figure \ref{fig;gr} and calculated
           PDFs for a structural model showing full JT distortions
           on all sites are shown as dotted lines. See text for more
           details.}
  \label{fig;fits}
\end{figure}

Local structural deviations from the average structure will show up as
deficiencies in the agreement since the fits were constrained so the
model has the average structure. We are particularly interested in
the size and shape of the {\it local} MnO$_6$ octahedron; we therefore
concentrate on the low-$r$ region of the PDF.
An enlarged
view of the region around the nearest neighbor Mn-O distance is shown in
Figure \ref{fig;fits}. 
The experimental data are shown as open circles. Two
model PDFs are
shown: The solid line represents the PDF of the refined average structural
model for doped manganites. Although the average structure is orthorhombic, the difference
in the three distinct Mn-O bond-lengths is very small making the
MnO$_6$ octahedra virtually regular.  The dotted
line is the PDF calculated from the average structure of undoped
LaMnO$_{3}$ where all Mn-O octahedra have a large 
JT distortion, i.e. short and long Mn-O bonds are present.  These are
clearly resolved in the 
calculation.\cite{proff;prb99;unpub} All the model curves are convoluted
with the experimental real-space resolution function of the data
which comes from the finite $Q$-range of the data.

It is apparent from Fig.~\ref{fig;fits}
that the model based on the average structure fits the $x=0.33$ and 0.25
data quite well in this low-$r$ region
but less well in the $x=0.12$ data set.  In fact, in the $x=0.12$ data
the dashed line representing the
JT distorted octahedra does a qualitatively better job of reproducing the
shape of the Mn-O bonds in the region 
$2.0 - 2.5$~\AA\ and the shape of the second
neighbor multiplet around 2.4 -- 2.8~\AA .  This supports the idea that,
locally, large JT distortions persist 
in the  insulating phase although these do not show up 
in the average crystal structure. In the ferromagnetic
metallic phase ($x=0.25$ and 0.33) the local structure is much 
closer to the average crystal structure.


\subsection{Low temperature structure of the MnO$_6$ octahedra}

\begin{figure}[!tb]
  \centering
  \includegraphics[angle=0,width=2.8in]{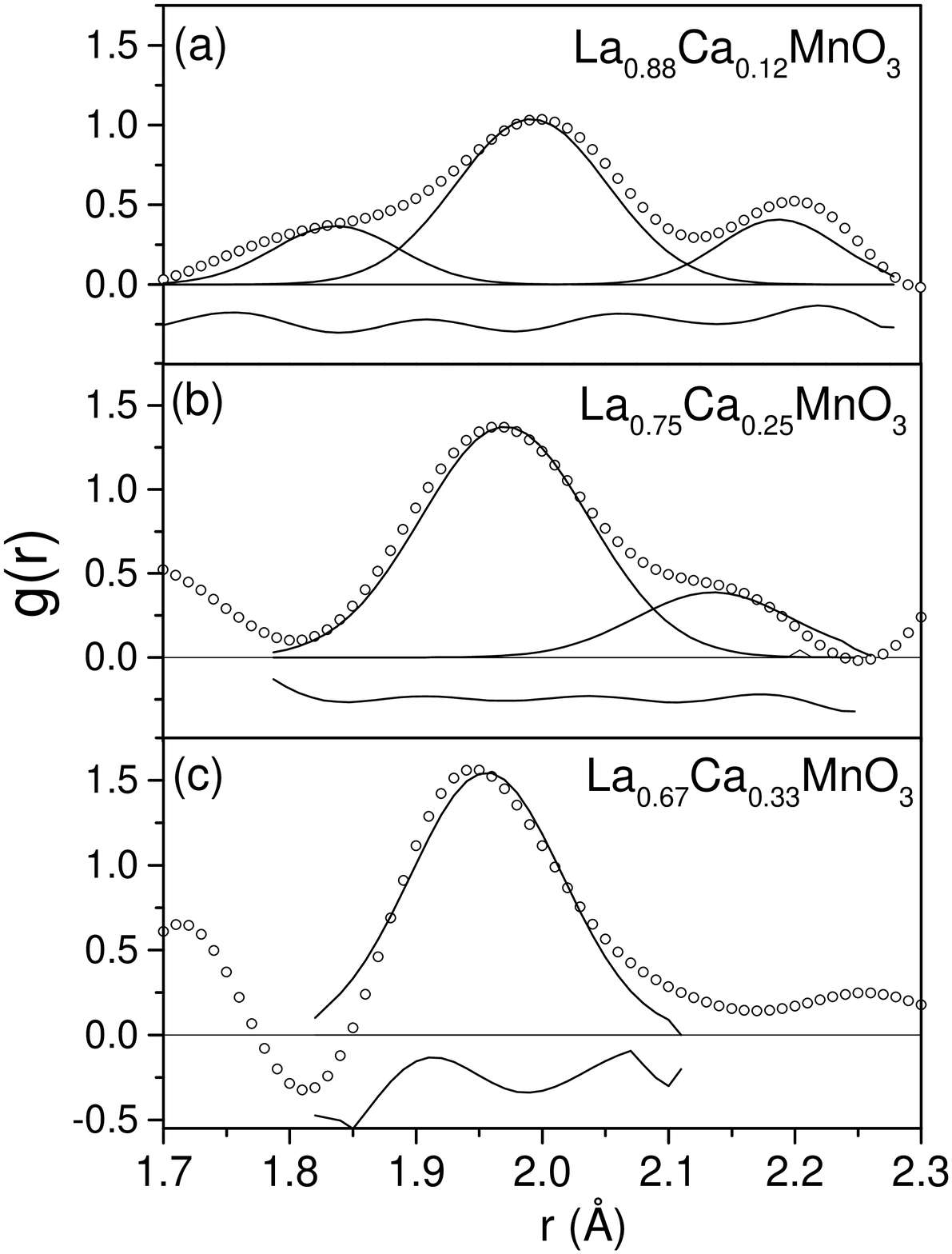}
  \caption{Low-$r$ region of the PDF on an expanded scale from
    $x=0.12$, 0.25 and 0.33 at 20~K (open circles).  Solid lines are 
    Gaussian fits to the data. The function $g(r)$ which
           is plotted is defined as $g(r) = 1 + {G(r)\over 4\pi r \rho_0}$ where
           $\rho_0$ is the number density of the material. 
  }
  \label{fig;lowrvx}
\end{figure}

We now focus on the region of the PDF from
$1.7\le r\le 2.3$~\AA\ containing the peaks from the MnO$_6$ octahedra.
This region is shown on an expanded scale in Fig.~\ref{fig;lowrvx} for 
doping levels $x=0.12$, 0.25 and 0.33 at 20~K.

We are interested to know how the MnO$_6$ octahedron 
evolves as a function of doping.  At $x=0.33$ the first PDF peak is
fit with a single Gaussian. There is a 
suggestion of peak asymmetry, but there is negligible intensity in the
region above 2.1~\AA .  
This single Gaussian fit means that all six Mn-O bonds have almost the same length of $r=1.96$~\AA\
at $x=0.33$ and $T=20$~K.  This is what would be expected for a fully
delocalized charge state.

The PDF for $x=0.25$ sample clearly has intensity on the high-$r$ side of
its first peak at 1.95~\AA\ which has been fit with a second Gaussian
component.  The presence of intensity at this $r=2.15$~\AA\ position
remains invariant as $Q_{max}$ is varied (although the resolution of the
feature changes).  This suggests that it is real
and not artificial since noise artifacts 
and termination ripples change position and intensity as $Q_{max}$
is varied. The suggestion is, therefore, that even at $T=20$~K
and $x=0.25$ long Mn-O bonds and, therefore, residual Jahn-Teller distorted sites 
persist in the material.  There is no direct evidence for intensity on the
low-$r$ side of the main 1.95~\AA\ peak although it does not decrease
as sharply as the $x=0.33$ sample. 

The $x=0.12$ sample is in the insulating state and is expected to be
fully localized and polaronic.  In this case we see three components
to the peak and have fit it with three Gaussians.  
At this composition there exist nominally Mn$^{3+}$ octahedra
which are Jahn-Teller distorted.  Based on the structure of undoped
LaMnO$_{3}$, we expect these to have four short bonds at 1.92--1.96~\AA\ and
two long bonds at 2.18~\AA . The two higher-$r$ components of the peak
seem consistent with this allocation.  The third component at low-$r$ might 
then be expected to originate from the Mn$^{4+}$ polaronic sites.
This is consistent with the prediction of the breathing mode 
model\cite{billi;prl96} which suggests short polaronic bonds of
$<1.9$~\AA , and is also consistent with the crystal chemistry of 
Mn$^{4+}$. Based on Shannon's ionic radii\cite{shann;ac76} the
expected Mn$^{4+}$(VI)-O$^{2-}$(II) ionic radius is 1.88~\AA.
This short Mn-O bond length is also found in the material 
CaMnO$_{3}$\cite{poepp;jssc82}
where all Mn sites are nominally Mn$^{4+}$.
It appears clear that Mn$^{4+}$ polarons exist with bonds 
$\le 1.9$~\AA\ together with Mn$^{3+}$ sites which have a JT 
distortion which is similar to that in the undoped material.


\subsection{Temperature dependence of the MnO$_6$ octahedra}

\begin{figure}[!tb]
  \centering
  \includegraphics[angle=0,width=2.8in]{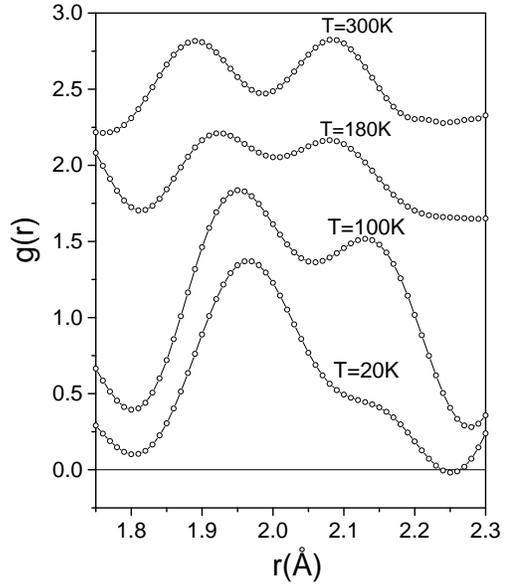}
  \caption{Low-$r$ region of the PDF on an expanded scale from
    $x=0.25$ at 20, 100, 180 and 300~K.  The data-sets are offset for
    clarity.
  }
  \label{fig;lowrvt}
\end{figure}

We now concentrate on the temperature evolution of MnO$_6$ octahedra.
In Fig.~\ref{fig;lowrvt} we show the evolution of the PDF peaks around
$r=2.0$~\AA\ as a function of temperature between $T=20$~K and
$T=300$~K for the $x=0.25$ sample. The metal-insulator transition
for this sample is at $T=235$~K.
The data-sets are offset for clarity.  As we discussed in the 
previous section, at low temperature a large central peak
centered around 1.97~\AA\ is evident with a small high-$r$
component at 2.15~\AA . 
As temperature is raised, the intensity 
in the high-$r$ component increases.  It is dangerous to 
infer a bond length directly from the position of a maximum in the data
because of the influence of noise on these data.  The small intensity
of these peaks is evident in Fig.~\ref{fig;gr} and  noise contamination
can cause a peak intensity to be shifted somewhat.  However, the 
presence or absence of intensity at some position is a robust result.
It is clear that the intensity of the peak grows in the region
$r=2.1$--2.15~\AA\ and below 1.9~\AA . It is also apparent that there is
no intensity above 2.2~\AA\ suggesting that the JT long-bond is 
$\sim 2.15-2.18$~\AA .

\begin{figure}[!tb]
  \centering
  \includegraphics[angle=0,width=2.8in]{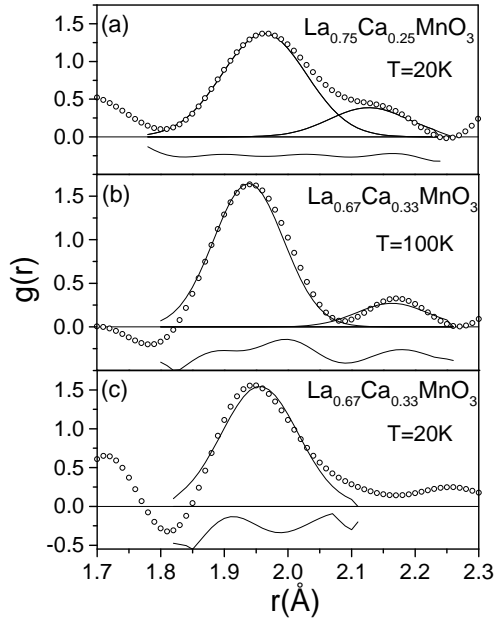}
  \caption{Low-$r$ region of the PDF on an expanded scale from
    $x=0.33$ at 20~K and 100~K and from $x=0.25$ at 20~K.
    The data-sets are offset for
    clarity.
  }
  \label{fig;lowrvtx}
\end{figure}

It is also interesting to see how the PDF of the MnO$_6$ octahedron
of the $x=0.33$ sample evolves with temperature.  This is shown in 
Fig.~\ref{fig;lowrvtx}.
At low temperature this sample exhibited a single peak centered
at $\sim 1.96$~\AA . There is no evidence of any JT long-bond.
At 100~K clear evidence of a component of intensity at $r=2.18$~\AA\
appears. The central peak also comes down less steeply on the low-$r$ side
and the peak centroid is shifted somewhat to lower-$r$ which suggest
that some intensity is appearing on the low-$r$ side of the peak.

It is interesting to note the similarity between the $x=0.33$ sample
at 100~K with the $x=0.25$ sample at 20~K which is shown in the top
panel of Fig.~\ref{fig;lowrvtx}.


\section{Discussion}

The ability to collect high-quality data at high values of the wavevector, $Q$, 
using high-energy synchrotron radiation has allowed us to decompose the bond 
length distribution of the MnO$_6$ octahedra into its components more reliably than 
was previously possible with neutrons. This is well demonstrated by the fact that the
present PDF data are consistent with the structure models derived by 
independent Rietveld studies and, furthermore, produce physically 
reasonable values for
the short Mn$^{4+}$ and the short and long Mn$^{3+}$ bonds. 

We draw the following conclusions from our
results described above.  If we assume that the $x=0.12$ sample is 
fully localized in the polaronic state at 10~K, as is suggested by
by its exponential resistivity~\cite{billi;prl96,jonke;p50,jaime;prb96}, 
we can 
interpret the three components of the first peak in the PDF as being due to
JT distorted Mn$^{3+}$  octahedra and regular but contracted
Mn$^{4+}$ octahedra.  If we assume the number of doped holes, $p$, to be
the nominal Ca concentration, $x$, then we expect the number of 1.88~\AA\
bonds to be $6p=6x=0.72$.  The number of JT distorted sites will be
$(1-p) = (1-x)$.  Then, if we assume that the JT distorted Mn$^{3+}$
octahedra have
essentially 4 short and 2 long bonds with average lengths 1.95 and
2.18~\AA , as observed in the undoped 
material,~\cite{proff;prb99;unpub,elema;jssc71} we expect a peak
with intensity $4(1-x) = 3.52$ at 1.95~\AA\ and $2(1-x)=1.76$ at
2.18~\AA .  Fitting the first peak in the experimental PDF with Gaussians 
(Fig.~\ref{fig;lowrvx}(a)) yields 
subcomponents with intensity ratios of 1.0(5):4.0(5):1.0(5) 
centered at 1.84, 1.96 and
2.18~\AA\ (the corresponding values for the model are
0.72:3.52:1.76 at 1.88, 1.95 and 2.18~\AA ).  
Given the noise level of the data the agreement is satisfactory 
providing some confidence to this interpretation.

We would like now to expand on the interplay between the
polaron formation and the 
existence of JT distortions in the manganites studied.
A simple picture could be constructed as follows:  There are no
distorted Mn-O octahedral units in the delocalized Zener phase and all Mn-O
bond-lengths in the MnO$_{6}$ octahedron are $\sim 1.97$~\AA , as found in the
average crystal structure. This is the case observed at x=0.33 and T= 20 K so that the
sample may be considered to be in fully delocalized charge state.
In the insulating phase there coexist
small, regular Mn$^{4+}$ octahedra with six Mn-O bonds of $\sim 1.85- 1.9$~\AA\
and Jahn-Teller distorted Mn$^{3+}$ octahedra with four bonds of $\sim 1.97$~\AA\ 
and two bonds of $\sim 2.18$~\AA\
length. This is the picture which we see at $x=0.12$, $T = 20$~K 
and at $x = 25$, $T = 300$~K (see Figs.~\ref{fig;lowrvx}(a) and
\ref{fig;lowrvt}). These two samples are  
in the
 insulating phase and the charge carriers, as the measured Mn-O bond 
length distributions suggest, are essentially fully localized. 
Within the FM phase but at 
intermediate temperatures, and compositions approaching the MI transition,
we see evidence for JT-long bonds appearing.
This suggests that there is
a coexistence of localized Jahn-Teller phase and delocalized
Zener phase material.  The sample is still conducting because the regions of
Zener phase percolate.  This is similar to the picture emerging for the
MI transition in La$_{0.625-y}$Pr$_{y}$Ca$_{0.375}$MnO$_3$ which occurs
as a function of $x$,\cite{uehar;n99} although the length-scale of the
inhomogeneities is much smaller in this case.

\begin{figure}[!tb]
  \centering
  \includegraphics[angle=0,width=2.8in]{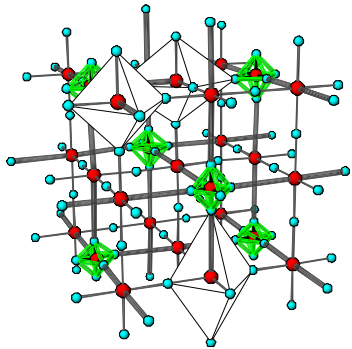}
  \caption{Ordered polaron model for $x=0.25$.  The small octahedra are on 
1/4 of the sites and contain localized Mn$^{4+}$ ions.  The other Mn sites
(large balls) have elongated JT distorted octahedral coordination
 whose long-bonds are
represented by thick lines.  
These can be oriented along $x$, $y$ or $z$
and always point towards a polaron.  Three of the octahedra have been drawn in
to illustrate this fact. For clarity
the magnitude of the polaronic/Jahn-Teller
distortions are exaggerated.
  }
\label{fig;distmod}
\end{figure}

Our picture is consistent with the earlier observation of a 
breathing mode distortion on one-in-four manganese sites which set
in below the MI transition in La$_{0.79}$Ca$_{0.21}$MnO$_{3}$.\cite{billi;prl96}
This was found to reproduce the changes in the local structure which occur at the
MI transition in this sample when the amplitude of the collapse was 
$\delta=0.12$~\AA .
Since the starting value of the Mn-O bond-length at low temperature
before the distortion set in was 1.97~\AA\ this results in short Mn$^{4+}$
bonds of 1.85~\AA\ shorter than, but similar to,
 what we observe here.  Furthermore, because
the model was evaluated at the special composition of $x=0.25$, this
breathing mode collapse coincidentally 
resulted in Jahn-Teller-like distortions
on {\it all} the remaining Mn sites with Mn$^{3+}$-long bonds of 
$1.97+0.12=2.09$~\AA .  This is 
illustrated schematically in Fig.~\ref{fig;distmod}.
This model gives a very satisfactory agreement with the
current results given
its simplicity.  
We do not wish to imply here that the
breathing mode collapse causes the Jahn-Teller distortion;
merely that they coexist in the localized phase and that there is
good consistency between the earlier neutron data and the current
x-ray data.  
\begin{figure}[!tb]
  \centering
  \includegraphics[angle=0,width=2.8in]{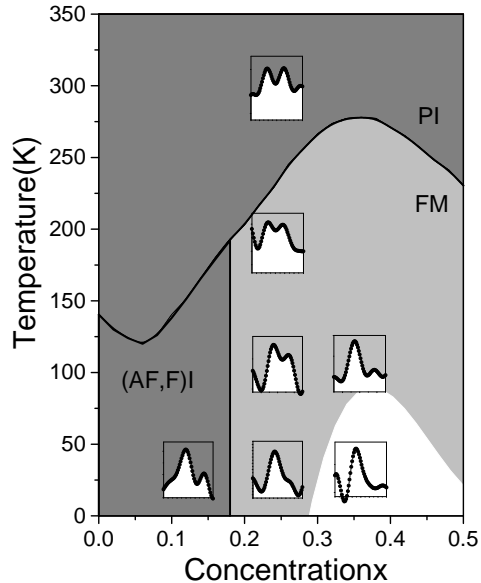}
  \caption{Schematic phase diagram for La$_{1-x}$Ca$_{x}$MnO$_{3}$.  The
solid lines are electronic and magnetic transitions taken from 
Refs.~\protect\onlinecite{cheon;b;cmarp99} and 
\protect\onlinecite{ramir;prl96}.  ``I'' and ``M'' refer to 
insulating and metallic respectively and ``P'', ``F'' and ``AF'' to
paramagnetic, ferromagnetic and antiferromagnetic.  Superimposed on
the figure are the low-$r$ PDF peaks showing the MnO$_6$ octahedra.
In the insulating phases the JT long-bond is clearly apparent.
There is no JT-long bond deep in the FM phase; however, it gradually
appears as the MI boundary is approached.  The fully localized polaronic
phases have dark shading.  The light shading indicates the region where
localized and delocalized phases coexist and the white region indicates
homogeneous FM phase.  The boundaries between the shaded regions are
diffuse and continuous and are meant to be suggestive only.
  }
  \label{fig;phased}
\end{figure}

It is interesting to note from Fig.~\ref{fig;distmod}
that $x=0.25$ is a special composition where
small polarons can form an ordered lattice separated by JT distorted
Mn$^{3+}$ sites which are unstrained.  Each Mn$^{4+}$ site has 6 neighboring
Mn$^{3+}$ sites whose long-bonds point towards it and these complexes fit
together into a space filling 3d network. There is no experimental
evidence that polarons order in this way in this system; 
rather charge stripes are observed.\cite{cheon;b;cmarp99}
However, this model does show how orbitals can order locally
around an Mn$^{4+}$ defect site to minimize strain.

So far, we have shown that by studying the size and shape of the MnO$_6$
octahedra we can determine whether the charge is localized as small
polarons (observation of 1.88~\AA\ and 2.18~\AA\ Mn-O bonds in the PDF)
or delocalized (observation of a single Mn-O bond length $\sim 1.97$~\AA).  
These two states are
exemplified by the $x=0.12$ sample at 20~K (Fig.~\ref{fig;lowrvx}(a))
and the $x=0.33$ sample at 20~K (Fig.~\ref{fig;lowrvtx}(c)) respectively.
As the temperature is increased below T$_c$ in 
the $x=0.33$ and $x=0.25$ samples, 
significant components of the long and short bonds become evident.
This suggests that carriers are becoming localized in parts of the sample .  The
high-resolution PDF data therefore support the idea of an inhomogeneous
sample with charge delocalized metallic regions of Zener phase coexisting with
regions of charge localized JT phase.  As T$_c$ is approached from below the
amount of charge localized phase increases at the expense of the charge delocalized
phase as evidenced by the growth of intensity at approximately 1.88 
and 2.1~\AA\ on increasing temperature in the $x=0.25$ 
sample (Fig.~\ref{fig;lowrvt}).
This view is consistent with Booth {\it et al.}'s
interpretation of their XAFS data\cite{booth;prb96,booth;prb98} 
and the interpretation
of our earlier neutron PDF data.\cite{billi;prl96,billi;b;pom99}
The MI transition and the onset of long-range ferromagnetic order
presumably coincides with the percolation of the Zener phase.
This is similar to the original proposition that the MI transition was
a percolation transition by Louca {\it et al.},\cite{louca;prb99} 
though our data support
the idea that the Zener phase percolates rather than a network of 
connected 3-site polarons.  A number of theories predict
charge phase segregation\cite{yunok;prl98} 
or two-fluid behavior\cite{jaime;b;pom99} 
of the charge system.  We note that our data is entirely consistent with
the coexistence of delocalized Zener phase and localized JT phase below
T$_c$ but does not directly imply the existence of charge
segregation between these phases; rather it is just the state of 
localization of the charges which differs in the different regions of the
sample,\cite{billi;b;pom99} as is proposed for 
La$_{0.625-y}$Pr$_{y}$Ca$_{0.375}$MnO$_3$.\cite{uehar;n99} 

Finally, we address the issue of how the charge state evolves
as the MI transition is approached  as a function of doping $x
$. As the experimental data suggest for $x=0.33$ and T=20 K  no polaronic or JT
distortions are present i.e. the true ground-state of the FM metallic phase
is a completely delocalized Zener state. Although we have a sparse data-set
one can notice the similarity of the Mn-O bond length distributions for 
$x=0.33$ at 100~K and $x=0.25$ at 20~K. Thus, as one moves away from
the ground-state a localized state appears that coexists with the 
delocalized one.  The volume of the localized state increases
as the MI transition is approached, whether as a
function of temperature (see Fig. \ref{fig;lowrvt}) or doping (see Fig. 
\ref{fig;lowrvx}).
The MI transition itself 
then occurs when the proportion of delocalized phase
is too small to percolate.  This view is summarized in 
Fig.~\ref{fig;phased}. In this figure the first peaks in the experimental PDFs,
reflecting the  MnO$_6$ octahedral  bond length distribution,  are plotted
at the positions on the phase diagram where they were measured.
The dark shading signifies essentially fully charge localized material; the
light shaded areas indicate a coexistence of charge localized and delocalized
phases and the white area is the fully charge delocalized region.  The
positions of the MI transitions are taken from standard phase diagrams
of this system.\cite{cheon;b;cmarp99,ramir;prl96} 
 

\section{Conclusions}
Using the PDF analysis of high energy x-ray diffraction we can
distinguish the charge localized and charge delocalized states
 of La$_{1-x}$Ca$_{x}$MnO$_{3}$ ($x=0.12$, 0.25, 0.33).  We characterize the nature of the
polaronic distortion around Mn$^{4+}$ as being an isotropic octahedron
with Mn-O bond length of $\sim 1.88-1.9$~\AA .
The FM phase is only homogeneous at low temperature and high doping
above $\sim x=0.3$.  As temperature is raised or doping lowered towards
the MI transition the FM state becomes inhomogeneous with a coexistence
of localized JT phase and delocalized Zener phase.


\acknowledgements
We would like to thank J. D. Thompson and M. F. Hundley for
help in characterizing the samples, Matthias Gutmann 
for help with data collection and E. Bo\v zin for a critical
reading of the manuscript.
We would like to acknowledge stimulating discussions with
P. G. Radaelli.
This work was supported by the NSF through grant DMR-9700966 
and by the Alfred P. Sloan Foundation. CHESS is funded by 
NSF through grant DMR97-13424.




\begin{thebibliography}{10}
\bibitem{lcmonote699}
Note that for neutron scattering the Mn-O peak in the PDF is negative because
  of the negative scattering length of Mn. However, for x-rays this peak is
  positive.
\end{thebibliography}
\end{document}